\documentclass[fleqn,usenatbib]{mnras}

\usepackage{newtxtext,newtxmath}
\usepackage[T1]{fontenc}

\DeclareRobustCommand{\VAN}[3]{#2}
\let\VANthebibliography\thebibliography
\def\thebibliography{\DeclareRobustCommand{\VAN}[3]{##3}\VANthebibliography}

\usepackage{graphicx}	
\usepackage{amsmath}	
\usepackage[inkscapelatex=false]{svg}
\usepackage{xspace}
\usepackage{gensymb}

\definecolor{TodoColor}{RGB}{239, 0, 101}

\newcommand{\rhat}{\ensuremath{\hat{\mathbf{r}}}}

\newcommand{\vdel}{\ensuremath{\mathbf{\nabla}}}

\newcommand\dd{\mathrm{d}}

\newcommand\peggy{V374 Peg\xspace}

\def\EXT{pdf}

\title[ECM on \peggy]{Electron Cyclotron Maser Emission from Ejected Stellar Prominences on \peggy}

\author[C. E. Brasseur et al.]{
C. E. Brasseur,$^{1,2}$\thanks{E-mail: cbrasseur@lowell.edu}
M. M. Jardine,$^{1}$\thanks{E-mail: mmj@st-andrews.ac.uk}
S. Daley-Yates,$^{1}$
J. F. Donati$^{2}$
J. Morin$^{3}$
\\
$^{1}$SUPA, Schools of Mathematics and Physics and Astronomy, North Haugh, St Andrews, Fife, KY16 9SS, UK\\
$^{2}$Lowell Observatory, 1400 W. Mars Hill Rd. Flagstaff, AZ. 86001. USA\\
$^{3}$Université de Toulouse, CNRS, IRAP, 14 avenue Belin, 31400 Toulouse, France\\
$^{4}$Université de Montpellier, CNRS, LUPM, F-34095 Montpellier,
France
}

\date{Accepted XXX. Received YYY; in original form ZZZ}

\pubyear{\the\year{}}

\begin{document}
\label{firstpage}
\pagerange{\pageref{firstpage}--\pageref{lastpage}}
\maketitle

\begin{abstract}

We investigate a possible origin for bursty radio emission observed on the active M dwarf V374 Peg, combining data-driven magnetic field modelling with archival radio light curves. We examine whether stellar prominence ejection can plausibly account for the observed radio bursts that have been attributed to electron cyclotron maser (ECM) emission. Our analysis shows that ejected prominences can produce the required energy range to drive the emission, and that modelled ECM visibility exhibits a rotational phase dependence consistent with the limited observational data (four observed bursts). The results support prominence ejection as a viable mechanism for ECM generation on V374 Peg and motivate further observational campaigns to constrain this process.

\end{abstract}

\begin{keywords}
stars: coronae -- stars: individual: V374 Peg -- stars: magnetic field -- stars: low-mass
\end{keywords}



\section{Introduction}

This study investigates whether ejected stellar prominences {could produce} the bursty radio emissions observed on some M dwarfs, using the star \peggy as a test case. By modelling prominence ejection events and comparing their predicted signatures with observed radio light curves, we aim to evaluate the plausibility of this mechanism as a driver of the observed emission. {One theory for the origin of} this bursty emission {is} electron cyclotron maser (ECM) emission, so here we specifically model the ECM flux resulting from ejected prominences on \peggy.

Coherent radio bursts have been observed on a number of stars, such as AU Mic \citep{Bloot_2024A&A...682A.170B}, UV Ceti \citep{Zic_2019MNRAS.488..559Z}, HR 1099 \citep{Slee_2008PASA...25...94S}, and CU Vir \citep{Trigilio_2000A&A...362..281T,Ravi_2010MNRAS.408L..99R,Leto_2006A&A...458..831L,Kellett_2007mru..confE.106K}. In a 2001 study \citet{Callingham_2021NatAs...5.1233C} presented a collection 19 coherent radio burst detections across a population of M Dwarfs, showing the ubiquity of such radio emission on the M Dwarf main sequence. {Additionally,} characterizing the contribution to ECM emission from stellar activity as opposed to other sources is of interest when studying a variety of phenomena, e.g. when searching for evidence of star-planet interaction \citep{Pineda_2023NatAs...7..569P}.

Stellar prominences are cool, dense structures that are trapped within the stellar corona and can give direct insight into the magnetic field conditions above the surface of the star (e.g. \citet{Cameron_1989MNRAS.236...57C,Ferreira_2000MNRAS.316..647F,CollierCameron_2003EAS.....9..217C,Jardine_2019MNRAS.482.2853J}). Prominences form in the corona itself and are insulated from the surrounding material by magnetic fields. Ejected prominences carry away mass and angular momentum, potentially impacting the atmospheric evolution of both the star itself as well as any orbiting exoplanets \citep{Villarreal_2018MNRAS.475L..25V}. Prominences are fed by plasma rising along magnetic field lines from the dense chromosphere. When the prominence mass becomes more than the magnetic field can support, the prominence will either fall back to the surface (prominences below the co-rotation radius) or be centrifugally ejected (prominences above, {so-called ``slingshot prominences''}) \citep{Jardine_2019MNRAS.482.2853J}. Recent simulations have indicated the ubiquity of condensations (prominences) in the atmospheres of cool stars, and the persistence of such cool condensations, even through prominence ejections large and frequent enough to be an {important} mechanism of mass-loss for fast-rotating stars \citep{DaleyYates_2023MNRAS.526.1646D,DaleyYates_2024MNRAS.534..621D,Waugh_2021MNRAS.505.5104W}.

{We theorize that} ejected prominences may induce ECM emission, which can be observed at radio wavelengths specific to the local magnetic field strength at the emission site. Because ECM emission is highly directional, emission from such ejections would appear as sharp increases in luminosity at specific phases as the star rotates {\citep{Stupp_2000MNRAS.311..251S}}. Such bursts, with a variety of characteristics, have been seen in the radio emission from a large number of M Dwarfs, and questions remain about the mechanisms driving it \citep{Villadsen_2019ApJ...871..214V}. We know that \peggy hosts prominences, and produces bursty radio emission \citep{Vida_2016A&A...590A..11V,Villadsen_2019ApJ...871..214V}, thus it offers the opportunity for us to model the ejection of its prominences and determine if their ejection is a plausible source for the observed radio bursts. {Our goal with this study is not to rule out other sources of radio bursts, {such as gyrosynchrotron emission}, but to investigate if prominence ejection induced ECM emission is a possible mechanism.}

\section{Prior V374 Peg studies}\label{ch4:vpeg}

\peggy was first catalogued in 1995, and was soon discovered to flare strongly and often \citep{Reid_1995AJ....110.1838R,Greimel_1998IBVS.4652....1G}. In 2001, \citeauthor{Montes_2001MNRAS.328...45M} showed that \peggy was part of the Castor kinematic group, and were thus able to determine its age (200 Myr). \peggy is a fully convective M4 dwarf that that sits near the boundary between fully and partially convective stars \citep{Llama_2018ApJ...854....7L,Vida_2016A&A...590A..11V}. Table \ref{tbl:vpeg_props} lists \peggy's relevant stellar properties. 

\citet{Donati_2006Sci...311..633D} presented a Zeeman Doppler Imaging (ZDI) surface magnetic field map of \peggy, from which they determined that \peggy rotates as a solid body and has a strong, largely poloidal and axisymmetric magnetic field. \citet{Morin_2008MNRAS.384...77M} used observations of \peggy spanning more than a year and concluded that the magnetic field was stable on that timescale, primarily poloidal, axisymmetric, and simple. \citet{Vida_2016A&A...590A..11V} presented a much longer-term study and concluded that \peggy's magnetic field is stable on a time period of at least 16 years. Additionally, over that span, they found no evidence of an activity cycle. Both these studies also found only very weak differential rotation (shear on the order of 10\% that of the Sun), supporting earlier findings of near-rigid-body rotation \citep{Morin_2008MNRAS.384...77M,Vida_2016A&A...590A..11V}. {More recently \citet{Bellotti_2025arXiv251109312B} have analysed additional spectropolarimetric observations of \peggy and concluded that its magnetic topology is stable on the order of 14 years.}

\citet{Vidotto_2011MNRAS.412..351V} used magnetohydrodynamic simulations built on a surface ZDI map to explore \peggy's wind. Their models suggest low coronal base densities (and thus a lower heating rate) on the order of $\lesssim 10^{17}~\mathrm{m}^{-3}$ (in contrast to, for example, the active M Dwarf AD Leo, which has a base density of $\sim 10^{18}~\mathrm{m}^{-3}$ \citep{Maggio_2005ApJ...622L..57M}). They showed that the ram pressure of \peggy's wind is about five times that of the Sun, meaning that planets orbiting in \peggy's habitable zone would need magnetic fields at least half the strength of Jupiter's for their atmosphere to be shielded against erosion from the stellar wind.

\begin{table}
    \centering
    \begin{tabular}{ccc}
    \hline
        \textbf{Property} & \textbf{Value} & \textbf{Source} \\
       \hline
       Mass & $0.28 \pm 0.05 M_\odot$ & [1]\\ 
       Distance & 9.1 pc & [2]\\
       Rotation Period &  $0.4456 \pm 0.0002$ day & [3]\\
       Inclination & $70 \pm 10 \deg$ & [3]\\
       Radius & $0.34 R_\odot$ & [1]\\
       Co-rotation radius ($r_K$) & $4.72R_*$ & \\
       Age & 200 Myr & [4]\\
       \hline
        \multicolumn{3}{l}{[1] \citet{Morin_2008MNRAS.384...77M}} \\
        \multicolumn{3}{l}{[2] Gaia DR3 (\citet{Gaia_2023A&A...674A...1G}} \\
        \multicolumn{3}{l}{~~~~~~\citet{Gaia_2023A&A...674A..28F,Gaia_2016A&A...595A...1G})}\\
        \multicolumn{3}{l}{[3] \citet{Donati_2006Sci...311..633D}} \\
        \multicolumn{3}{l}{[4] \citet{Montes_2001MNRAS.328...45M}} \\
    \end{tabular}
    \caption{Stellar properties of \peggy. }
    \label{tbl:vpeg_props}
\end{table}

\citet{Hallinan_2009AIPC.1094..146H} presented a radio light curve of \peggy that exhibits several radio bursts in addition to a quiescent flux that varies on the period of the star. They presented evidence that both components of the emission are highly directive emission, {and suggested} ECM emission {as the mechanism}.

\citet{Llama_2018ApJ...854....7L} used ZDI-based corona modelling to show that the quiescent radio flux could be reproduced by ECM emission. The rotation of \peggy is known with high enough precision that \citet{Llama_2018ApJ...854....7L} could align the phase of their synthetic light curve with the radio observation. They explored the effect of offsetting the phases of the synthetic data and observations on the fit and found that the larger the offset, the less well the simulated data could reproduce the observations. This provides additional evidence that the emission is a result of the magnetic field structure. 

Because \citet{Llama_2018ApJ...854....7L} modelled steady state ECM emission, they were only able to account for the phase-modulated quiescent light curve. Here we use a similar model, but consider ECM emission specifically induced by ejected prominences, an inherently intermittent phenomenon. Thus, in contrast to \citet{Llama_2018ApJ...854....7L}, we aim to reproduce the radio bursts rather than the quiescent flux.

{\section{Prominence induced Electron Cyclotron Maser emission}}

ECM emission is a form of nonthermal coherent emission that results from electrons being accelerated along a magnetic field line
and gyrating around the line at the local cyclotron frequency (the angular frequency at which an electron will oscillate for a given magnetic field strength), and when that frequency is greater than the local plasma frequency (the charge density oscillation frequency for plasma at a given density), they emit coherently, approximately perpendicular to the magnetic field line \citep{Treumann_2006A&ARv..13..229T,Trigilio_2000A&A...362..281T}. 

\begin{figure}
\centering
	\includegraphics[width=\columnwidth]{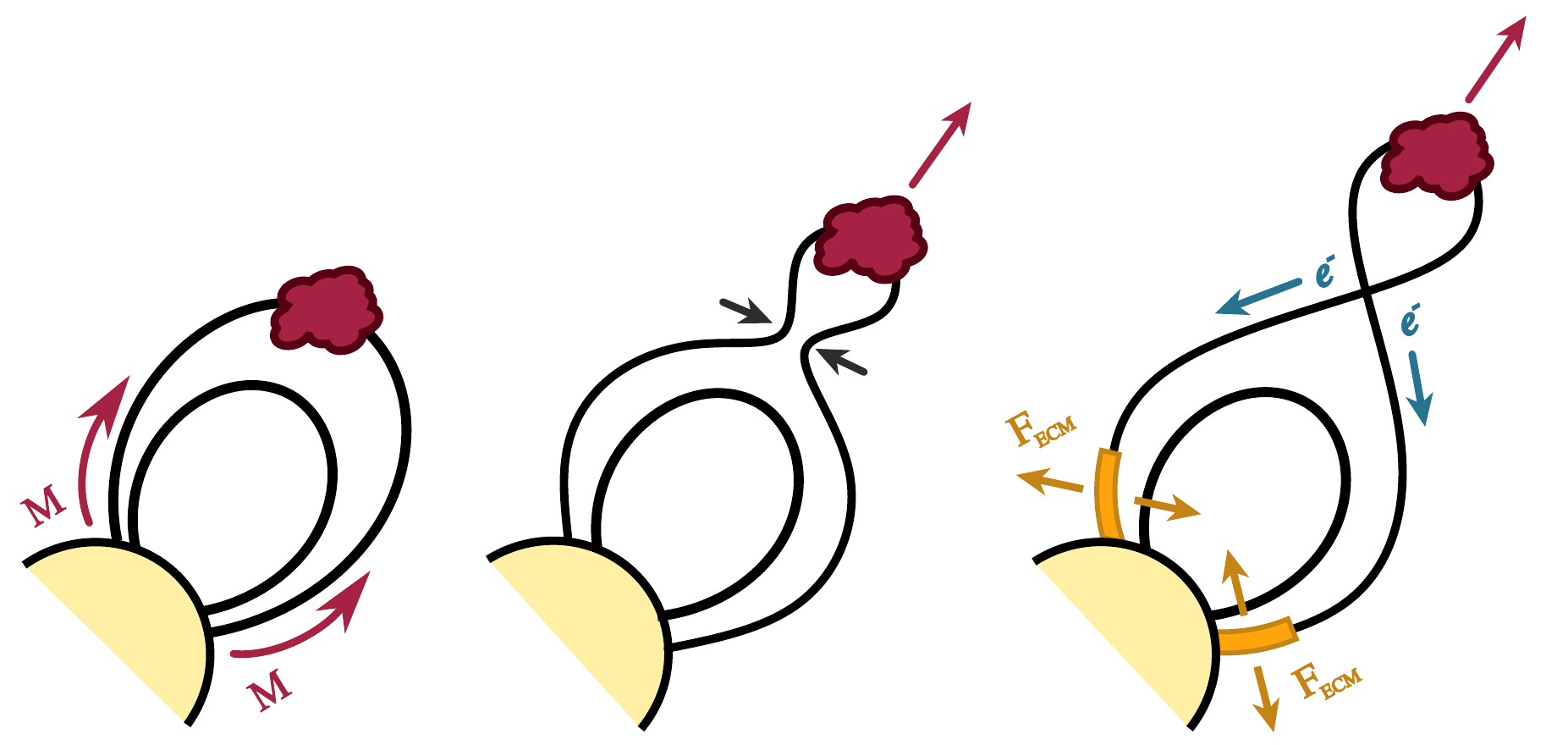}
    \caption{{Diagram illustrating how the ejection of a prominence at or beyond the co-rotation radius can power ECM emission. \textit{Left:} Mass flows up the field lines into the prominence. \textit{Middle:} The prominence becomes too massive for the magnetic field to support and begins to pull away, deforming the magnetic field lines as it does so. \textit{Right:} The prominence is ejected, causing a sudden reorganization of the magnetic field. This forces a beam of electrons down the magnetic flux tube, and induces ECM emission.}}
    \label{fig:ecm_exp}
\end{figure}

{The process linking prominence ejection to ECM emission is illustrated in Fig. \ref{fig:ecm_exp}. Mass flows upward along magnetic field lines into the prominence, gradually increasing its mass and distorting the magnetic structure. Once the prominence becomes too massive for magnetic support, if it is above the co-rotation radius, it begins to detach from the star, stretching and distorting the surrounding magnetic loops. Eventually, magnetic reconnection separates the loop supporting the prominence from the rest of the magnetic field, rapidly reorganizing the magnetic topology and releasing energy as electrons are accelerated along the field lines. These electrons develop a non-thermal energy distribution, which can drive the electron cyclotron maser instability and produce ECM emission \citep{Treumann_2006A&ARv..13..229T,Trigilio_2000A&A...362..281T}. } Here we are exploring ECM emission induced by ejected stellar prominences as a source for the radio bursts observed on \peggy.

\begin{figure}
	\includegraphics[width=\columnwidth]{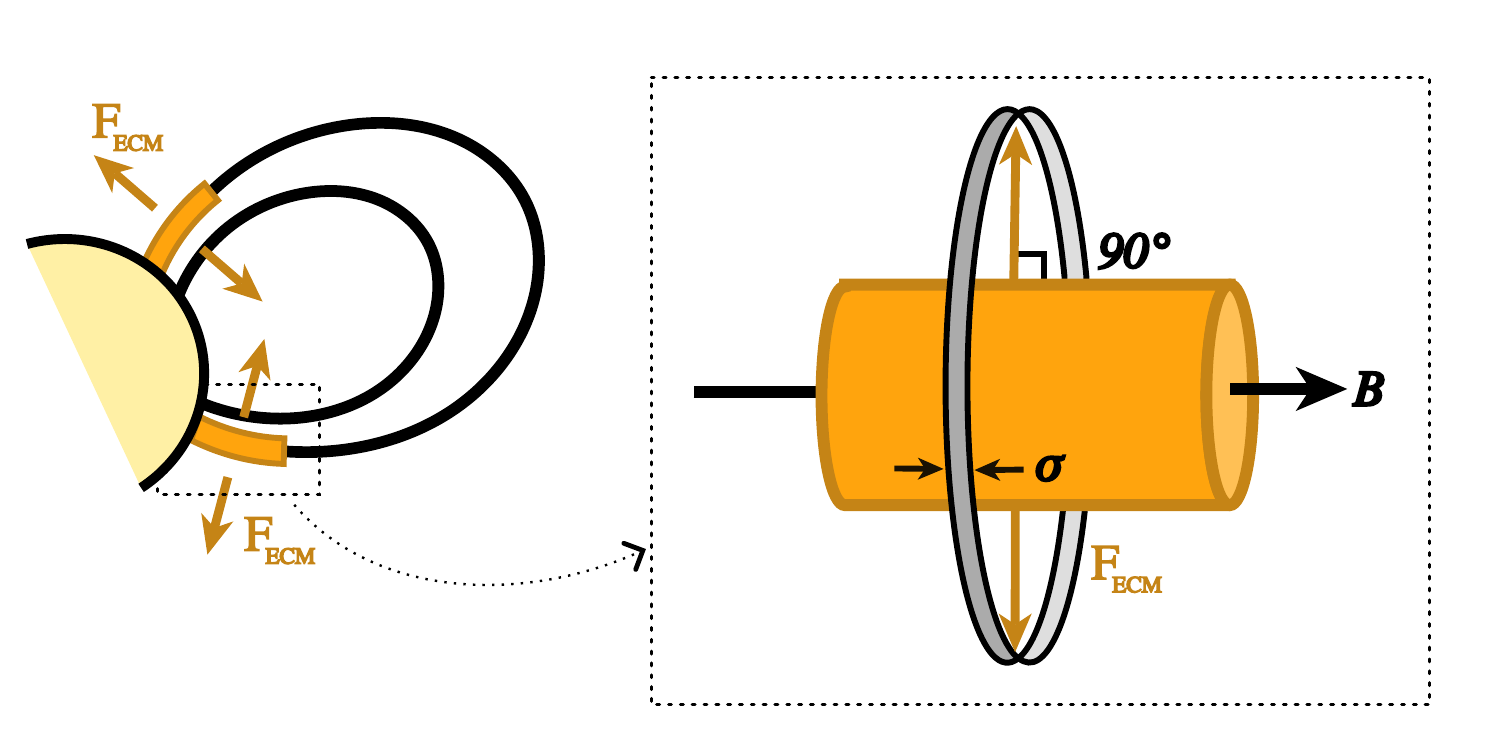}
    \caption{Diagram illustrating ECM emission {directionality. The parts of the flux tube where conditions allow ECM emission are those closest to the star. The photons are emitted perpendicular to the flux tube in a hollow cone with an opening angle of $90\degree$  and width $\sigma$ (see inset on right).}}
    \label{fig:ecm_diagram}
\end{figure}

{Because the origin of the ECM emission process is the prominence ejection,we use the gravitational potential energy of the prominence as an upper limit to the energy available for the ECM emission. However, none of the  energy transformations (prominence ejection, to magnetic reconnection, to ECM instability) are perfectly efficient, so the total energy available for ECM emission is actually the gravitational potential energy times an efficiency factor.}
The available energy is then spread out along the magnetic field line, and {induces ECM emission where the conditions that allow such emission are met.} Each emitting point on the field line emits at its local gyrofrequency, into a hollow cone with an opening angle of $90\degree$ {as shown in figure \ref{fig:ecm_diagram}} \citep{Melrose_1982ApJ...259..844M}. Experimental work has estimated the efficiency of the electron beam to radio flux mechanism to be about 1\% \citep{MacKinnon_1992A&A...256..613M,McConville_2008PPCF...50g4010M}. 

\section{Methods}\label{ch4:methods}

The Potential-Field Source-Surface (PFSS) model is a relatively simple method for modelling a coronal magnetic field between the surface of the star (or, more precisely, the base of the corona) and the so-called ``source surface.''  The ``source surface'' is the theoretical radius at which all magnetic field lines become radial. It is one way to quantify the extent of the stellar corona, which in this definition encompasses the part of the stellar atmosphere characterised by closed magnetic field lines. While this imaginary ``surface'' is not in reality spherical (see, e.g. the results from the Parker Solar Probe, \citet{Panasenco_2020ApJS..246...54P}), it is a useful construct. 

At the centre of this model is the assumption that the stellar magnetic field is ``potential,'' in other words, that the curl of the magnetic field is zero ($\vdel\times\mathbf{B}=0$). This means that the  magnetic field ($\mathbf{B}$) can be described in terms of a scalar potential ($\Psi$)
\begin{equation}
  -\vdel \Psi = \mathbf{B}. 
\end{equation}
The requirement that the magnetic field is divergence free ($\vdel\cdot\mathbf{B}=0$) reduces the problem to Laplace's equation
\begin{equation}
  \nabla^2\Psi=0 \label{eqn:laplace}
\end{equation}
which is then solved in terms of spherical harmonics.

The inner boundary of the model is the radial component of a surface magnetic field map, and the outer boundary is a geometric condition that requires all field lines to be radial at the source surface ($\mathbf{B_{ss}} = B_{ss}\rhat$).  The lack of inertial forces is a limitation of this model, however, it still compares favourably to more complete magnetohydrodynamic models for applications where the large-scale magnetic structure is paramount \citep{Riley_2006ApJ...653.1510R,Asvestari_2024ApJ...971...45A}. For example, when \citet{Jardine_2013MNRAS.431..528J} explored the effect of the non-potential component of the stellar surface magnetic field, they concluded that the non-potential nature of the field does not significantly affect the stellar wind. {Similarly, \citet{Asvestari_2024ApJ...971...45A} show that for the sun, a PFSS model with spherical source surface well approximates the large scale magnetic field detecable in the radio regime, despite its limitations.}

For the inner boundary condition, we use ZDI surface magnetic field maps. ZDI is a spectropolarimetric tomographic technique that uses time series spectropolarimetric observations over at least a full rotation period to recover the large-scale magnetic field strength and geometry of a star \citep{Semel_1989A&A...225..456S}.

\subsection{V374 Peg Model Parameters}\label{ch4:vpeg_find_parms}

\begin{figure}
\centering
	\includegraphics{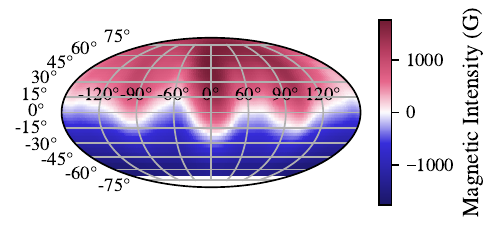}
    \caption[\peggy surface magnetic field map]{The radial component of the ZDI-determined surface magnetic field map for \peggy from August 2005 \citep{Donati_2006Sci...311..633D}.}
    \label{fig:surface_map}
\end{figure}

The model presented here is built from the 2005 ZDI surface magnetic field map of \peggy presented in \citet{Donati_2006Sci...311..633D} (Fig. \ref{fig:surface_map}). This is a ``maximum-entropy'' reconstruction, which minimises the scale-weighted magnetic energy as a sparsity constraint when building up the field topology to fit the observations. This map is not simultaneous with the radio light curve to which we will compare our model, however, as established above, \peggy has a magnetic field that is stable on the order of decades. Thus, the temporal difference of just over a year is not a concern for the fidelity of the comparison.

The free parameters in our model are: corona temperature, source surface radius, and base pressure. We will treat each of these in turn (the values we ultimately used are shown in Table \ref{tbl:vpeg_model}). Coronal temperature ($T_{cor}$) is the most straightforward. We adopt the value given in \citet{Llama_2018ApJ...854....7L}. They used X-ray flux measurements of \peggy to estimate $T_{cor} = 6\times10^6$~K, using the method outlined by \citet{Johnstone_2015A&A...578A.129J}.

The base pressure is the plasma pressure at the base of the corona, in our model given by 
\begin{equation}
    p_0(\theta,\phi)\mathrm{~[Pa]} = \kappa_p B_0(\theta,\phi)^2\mathrm{~[G]} \label{eqn:p0}
\end{equation}
where $B_0$ is the magnetic field strength on the surface and $\kappa_p$ is a scaling factor. We follow the method laid out in \citet{Jardine_2020MNRAS.491.4076J} to estimate the scaling factor $\kappa_p$ using the two constraints that the observed prominences must be magnetically confined and have an observed lifetime. This method involves calculating the plasma $\beta$ (the ratio between the plasma pressure and the magnetic pressure) at the co-rotation radius, and the prominence lifetimes, for a grid of pressure scaling factors and corona temperatures. 

In isothermal hydrostatic equilibrium, the plasma pressure at the equatorial co-rotation radius ($R_k$, where prominences are located) is expressed as
\begin{multline}
p_g(R_k) = p_0~\mathrm{exp} \Bigg\{  \frac{m}{k_B T_{cor}}\Bigg( \frac{GM_\star}{R_\star}\frac{R_\star-R_k}{R_k} +  \\ \frac{1}{2}\omega^2 (R_k-R_\star) (R_k+R_\star) \Bigg )  \Bigg\}\label{eqn:pgr}
\end{multline}
where $p_0$ is the pressure at the stellar surface, $m$ is the mean particle mass, $k_B$ is the Boltzmann constant, $T_{cor}$ is the corona temperature, $G$ is the gravitation constant, $M_\star$ and $R_\star$ are the stellar mass and radius respectively, and $\omega$ is the star's angular velocity.  We can similarly express the magnetic pressure as a function of radius with
\begin{equation}
    p_B(R_k) = \frac{B(R_k)^2}{2\mu} =  \frac{B_0^2}{2\mu} \left ( \frac{R_\star}{R_k} \right )^6\label{eqn:pbr}
\end{equation}
where $B_0$ is the magnetic field strength at the base of the magnetic field line passing through $R_k$, $\mu$ is the magnetic permeability, and all other variables are the same as in Eq. \ref{eqn:pgr} (see \citet{Brasseur_thesis} appendices C and D for derivations of these equations). Equations \ref{eqn:pgr}-\ref{eqn:pbr} define an expression for the plasma $\beta$, which, while complicated, depends only on $T_{cor}$ and $\kappa_p$.

\begin{figure}
    \centering
	\includegraphics{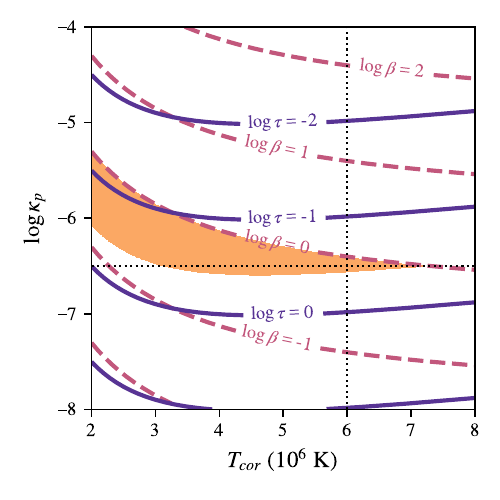}
    \caption[Prominence lifetime and plasma $\beta$ contours in temperature/pressure space]{Prominence lifetime ($\tau$) and plasma $\beta$ contours in coronal temperature ($T_{cor}$) vs pressure scaling factor ($\kappa_p$) space. $\tau$ is measured in days. The shaded region satisfies the criteria $\beta \le 1$ and $\tau \le 0.38$ days. The dotted lines mark $T_{cor} = 6\times10^6$ K and $\kappa_p = 10^{-6.5}$.}
    \label{fig:kappa_contours}
\end{figure}

{When estimating $\kappa_p$, we perform all calculations for a pure dipole, because this makes them analytically calculable. This is a reasonable simplification because in this calculation we are concerned with the magnetic pressure at the radius where prominences are found, at which radius the dipole component dominates. Additionally, in a pure dipole, because of the angular symmetry of the magnetic field geometry, all prominences have the same maximum mass and thus prominence lifetimes. For non-idealized magnetic fields with large dipole components (like that of \peggy), this dipole prominence lifetime is a good estimate of the average prominence lifetime.}

The lifetime of a prominence is given by $\tau = \frac{M_p}{\dot{M}}$ where $M_p$ is the maximum supported prominence mass and $\dot{M}$ is the mass upflow rate,  i.e. the rate at which mass is entering the prominence. The mass upflow rate can be calculated as
\begin{equation}
    \dot{M} = A \rho u
\end{equation}
where $A$ is the cross-sectional area of the flux  tube, $\rho$ is the mass density, and $u$ is the wind speed. Because $\dot{M}$  is constant along the flux tube, this can be calculated at any point on the tube, so for convenience, we calculate it at the foot point. We assume a thermal wind, so the wind speed at the surface of the star is only a function of the coronal temperature $T_{cor}$ \citep{Parker_1958ApJ...128..677P}. The maximum mass that can be supported can be found by balancing the centrifugal and magnetic tension forces 
\begin{equation}
 \rho g_{\mathrm{eff}}=\frac{B^2}{\mu R_{\mathrm{c}}},
\end{equation}
where $\rho$ is the prominence density,  $g_{\mathrm{eff}}$ is the local effective gravity, $R_c$ is the local radius of curvature, and $\mu$ is the magnetic permeability. Thus we can calculate a fiducial lifetime for a specific $T_{cor}$ and $\kappa_p$, and use that to calculate $\tau$ for the rest of the grid as
\begin{equation}
    \tau = \tau_0 \frac{\kappa_{p0}}{\kappa_p} \left (  \frac{T_{cor}}{T_0} \right )^{2.5} e^{2R_s(T_{cor}/T_0 - 1)}
\end{equation}
where $\tau_0$, $\kappa_{p0}$, and $T_0$ are the fiducial values, $R_s$ is the sonic point (the radius where the stellar wind speed is the sound speed), and all other variables are the same as above.

Fig. \ref{fig:kappa_contours} shows the contours through the $T$ vs. $\log  \kappa_p$ plane for $\tau$ and $\beta$. The shaded area represents values where the plasma $\beta$ is $\le1$ (a requirement for the prominences to be confined by the magnetic field), and the prominence ejection rate is at least 4 ejections in 36 hours, or $\tau = 0.38$ days (the number seen by \citet{Hallinan_2009AIPC.1094..146H}). Using the coronal temperature $T_{cor} = 6 \times 10^6~\rm{K}$ \citep{Llama_2018ApJ...854....7L} we determine that $\kappa_p = 10^{-6.5}$ (marked with the intersecting dotted lines). This corresponds to a base number density of $10^{15-16}~\mathrm{m}^{-3}$, which compares favourably to the finding of $\lesssim 10^{17}~\mathrm{m}^{-3}$ from \citet{Vidotto_2011MNRAS.412..351V}.

\begin{figure}
\centering
	\includegraphics{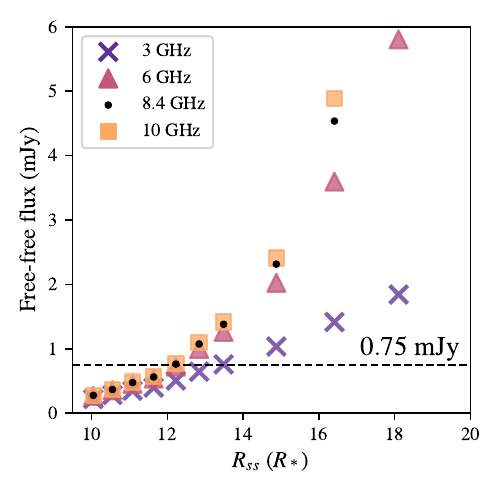}
    \caption{Free-free flux vs. source surface radius ($R_{ss}$) for \peggy. Frequencies that correspond to the {centres of the} VLA S-, C-, and X-bands, {as well as the observing frequency used by \citet{Hallinan_2009AIPC.1094..146H}} are calculated using the methods outlined in \citet{Brasseur_2024MNRAS.530.2442B}. The flux we fit the model to (0.75 mJy) is marked.}
    \label{fig:ff_by_rss}
\end{figure}

This leaves a single free parameter, the source surface radius $R_{ss}$. To determine that value, we fit the output of our model to measurements of \peggy's quiescent radio flux. \citet{Hallinan_2009AIPC.1094..146H} observed \peggy on three successive nights in January 2007, {centred at 8.4 GHz using the VLA X-band (standard bandwidth of 50 MHz)}\footnote{{The specific observing frequency is not mentioned in \citet{Hallinan_2009AIPC.1094..146H}, however we obtained the original observations from the VLA archive (under project code AH934), and were able to verify the observing band.}}. They observed bursty emission as well as quiescent background flux. 
{The observed quiescent flux includes all sources of quiescent radiation, not just free-free emission, and indeed \citet{Hallinan_2009AIPC.1094..146H} suggest that much of the quiescent flux is the result of highly directive emission, which free-free is not. This is born out by the fact that, for our range of models, the free-free emission modulation is an order of magnitude lower than the observed flux variation.} We therefore chose to fit the free-free emission to {0.75 mJy, which is the centre point of the range} observed by \citet{Hallinan_2009AIPC.1094..146H} ( $\sim 0.5-1$ {mJy}).

For a range of $R_{ss}$ values, we use the method described in \citet{Brasseur_2024MNRAS.530.2442B} to calculate the free-free flux at an arbitrary phase. The phase chosen does not affect the results because, for this range of models, the free-free flux modulation is very low as the star rotates. Fig. \ref{fig:ff_by_rss} shows the results of this exercise. {In this plot we show the free-free flux for the centre frequency of the VLA S-, C-, and X-bands, as well as the specific frequency at which \citet{Hallinan_2009AIPC.1094..146H} observed. Fitting the black (8.4 GHz) points, we} chose $R_{ss} = 12.2 R_\star$. 

\begin{table}
    \centering
    \begin{tabular}{rl}
        \textbf{Property} & \textbf{Value}  \\
       \hline
        ZDI map year$^{a}$  & 2005\\
        Corona temperature ($T_{cor}$)$^{b}$ & $6 \times 10^6~\mathrm{K}$\\
        Pressure scaling constant ($\log{\kappa_p}$) & -6.5\\
        Source surface radius ($R_{ss}$) & $12.2 R_*$\\
        Total efficiency factor & $10^{-6}$ \\
       \hline
       \multicolumn{2}{l}{$^{a}$ \citet{Donati_2006Sci...311..633D} } \\
 \multicolumn{2}{l}{$^{b}$ \citet{Llama_2018ApJ...854....7L}} 
    \end{tabular}
    \caption[\peggy model parameters]{\peggy model parameters used throughout this chapter.}
    \label{tbl:vpeg_model}
\end{table}

\begin{figure*}
\centering
	\includegraphics{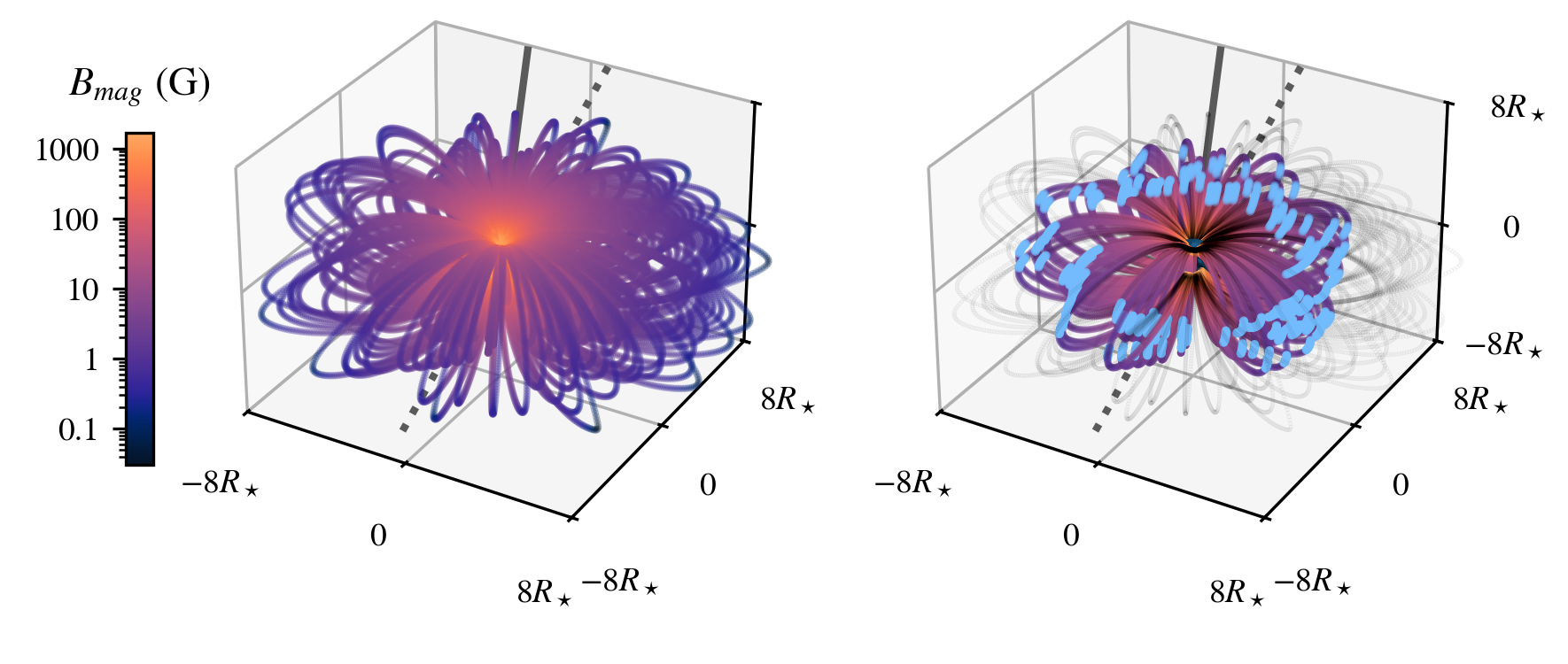}
    \caption[\peggy model magnetic field lines]{\peggy model magnetic field lines, given the parameters in Table \ref{tbl:vpeg_model}. The \textit{left} plot shows a representative sample of the closed field lines coloured by magnetic field strength. In the \textit{right} plot, only the prominence bearing field lines are coloured by magnetic field strength, and the prominences themselves are coloured blue. The rotation axis is marked with the dotted grey line, while the magnetic axis is solid. The magnetic axis is misaligned from the rotation axis by about $13\degree$.}
    \label{fig:vpfield_lines}
\end{figure*}

Fig. \ref{fig:vpfield_lines} shows the field line extrapolation for \peggy with the input parameters shown in Table \ref{tbl:vpeg_model}. The field lines are traced through the three-dimensional grid of the PFSS model described at the start of this section, which in this case is linear in $\theta$ and $\phi$, and exponential in radius. We can see that it has a very dipolar magnetic field with prominences arranged in a ring over the equator. The rotation axis is tilted by $70\degree$ from our viewing angle, and the magnetic axis is tilted a further $13\degree$, so observations of \peggy are nearly edge on to the magnetic equator \citep{Donati_2006Sci...311..633D}. 

\subsection{Synthetic Observations} 

Because ECM emission occurs at very specific frequencies that vary along the magnetic field line, the synthetic data product we chose to build is a dynamic spectrum (intensity vs. frequency vs. time), which can then be collapsed or sliced in the time direction to obtain a spectral energy distribution (SED), and in the frequency direction to obtain a light curve. 

\subsubsection{ECM flux calculation}\label{ch5:ecm}

The basic method for producing a synthetic dynamic spectrum is to calculate, for each magnetic field line with an ejected prominence, all the cells in which ECM emission occurs, and record the flux and frequency of the emission. Then the star is rotated, and the visible emission at each observing frequency is calculated at each viewing angle.

The condition for ECM emission to be possible is that the electron plasma frequency,
\begin{equation}
\omega_p=\frac{e}{2 \pi} \sqrt{\frac{n_e}{m_e \epsilon_0}} \approx 9 \sqrt{n_e\mathrm{~[m^{-3}]}} \mathrm{~kHz}
\end{equation}
must be less than the electron-cyclotron frequency,
\begin{equation}
    \Omega_e=\frac{e B}{2 \pi m_e} \approx 2.8 \times 10^6 B\mathrm{[G]} \mathrm{~Hz},
\end{equation}
where $e$ is the elementary charge, $n_e$ is the electron number density, $m_e$ is the electron mass, $\epsilon_0$ is the permittivity of free space, and $B$ is the magnetic field strength \citep{Treumann_2006A&ARv..13..229T}.

The energy powering the ECM emission results from the ejection event increasing the potential energy of the prominence rising towards zero as it is ejected, allowing the accelerated electrons to tap into the associated gravitational potential energy reservoir. Thus, the total energy available ($E_{tot}$) is the original gravitational potential of the prominence. However this is, of course, not a perfectly efficient process, so we must multiply the gravitational potential by an efficiency factor $\epsilon$. The energy is being emitted in the form of photons, meaning that we describe the emitted energy as a sum of the energies of the emitted photons. Putting this together, we can describe the emission with the following equations:
\begin{equation}
\begin{aligned}
    E_{tot} &= \epsilon \frac{G M_\star m_{prom}}{r_{prom}}\\
            &= \sum_i h \nu_i
\end{aligned}
\end{equation}
where $M_\star$ and $m_{prom}$ are the stellar and ejected prominence masses respectively, $r_{prom}$ is the distance between the centre of the star and the ejected prominence, $h$ is Planck's constant, and $\nu_i$ is the frequency of an individual photon.

\begin{figure}
\centering
	\includegraphics[width=.7\columnwidth]{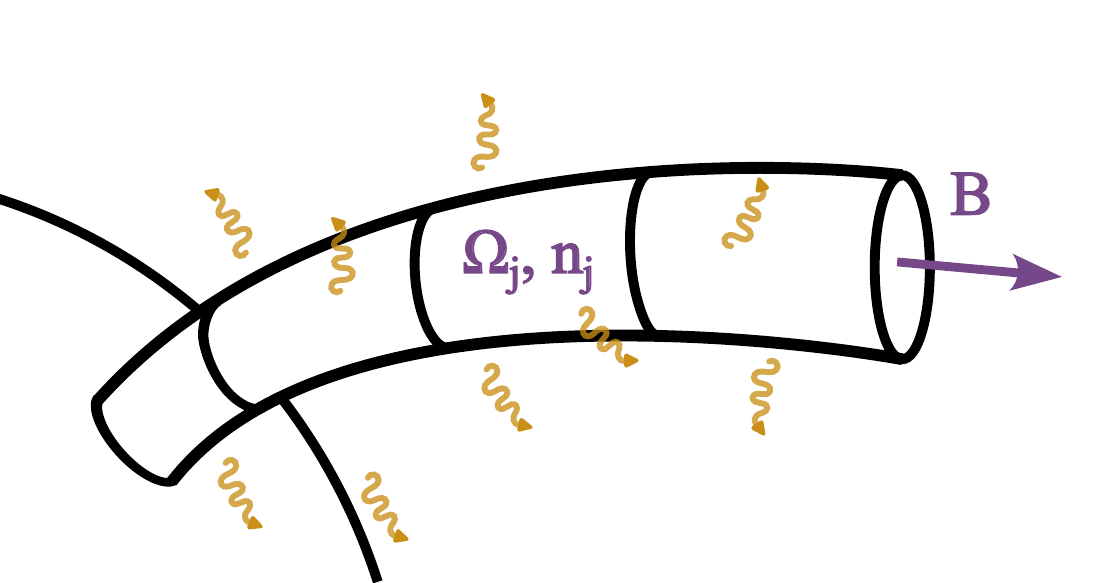}
    \caption[ECM emitting flux tube diagram]{Diagram of an ECM emitting flux tube divided into segments, each with a single number density ($n$) and gyrofrequency ($\Omega$).}
    \label{fig5:ecm_segments}
\end{figure}

In our model, the flux tube is divided into segments where each segment has a single value of magnetic field strength and density (Fig. \ref{fig5:ecm_segments}). We can thus calculate the emitted energy for each segment individually and take the sum for the total energy. We can thus write the segment energy as
\begin{equation}
    E_{j} = n_j h \Omega_j
\end{equation}
where $n_j$ is the number of photons in the segment, and $\Omega_j$ is the local gyrofrequency. The total energy is simply the sum of the energies from each individual segment
\begin{equation}
    E_{tot}=\sum_{j} E_{j} = \sum_{j} n_{j} h \Omega_{j}.
\end{equation}
We can therefore write the energy from each segment as a fraction of the total energy available
\begin{equation}
    E_{j} = \frac{h n_{j} \Omega_{j}}{\sum\limits_{j} h n_{j} \Omega_{j}} E_{tot} = \frac{n_{j} \Omega_{j}}{\sum\limits_{j} n_{j} \Omega_{j}} E_{tot}.
\end{equation}
Because the number of electrons per second passing through a given section of the flux tube is constant, the number density must be proportional to the segment length ($n_i \propto \dd s_i$), so we can write the segment energy as
\begin{equation}
    E_{j} = \frac{\dd s_{j} \Omega_{j}}{\sum\limits_{j} \dd s_{j} \Omega_{j}} E_{tot}
\end{equation}
This is now in a form that can be modelled. However, not every segment will actually be able to emit. This is because when the local plasma frequency is greater than the gyrofrequency, ECM emission is suppressed \citep{Treumann_2006A&ARv..13..229T}, so while the energy from the ejected prominence is spread along the flux tube, ECM emission will only be produced where conditions allow it to escape.

This energy is not emitted instantaneously, of course, and we model it as emitted over some timescale $t_{ej}$, meaning we can write the power from an emitting segment as
\begin{equation}
    P_{j}=\frac{E_j}{t_{ej}}
\end{equation}
and, given an observed frequency range $\Delta\nu$, the luminosity as
\begin{equation}
    L_j=\frac{(E_j/t_{ej})}{\Delta\nu}.
\end{equation}

To calculate the observed ECM flux, we need to know the surface area into which the segment is emitting. ECM emission is directional, with photons being emitted into a hollow cone distribution with angular thickness $\sigma\degree$ and opening angle of $90\degree$ (see Fig. \ref{fig:ecm_diagram}) relative to the magnetic field vector ($\mathbf{B}$). Experiments have shown that $\sigma$ is about $1-2\degree$, so in this work we have set $\sigma=1\degree$ \citep{Melrose_1982ApJ...259..844M}. Thus, the magnetic field segment emits into a cylinder with a radius of the distance to the observer, and height defined by the thickness angle $\sigma$ (see Fig. \ref{fig:ecm_diagram}). This means that the surface area of the cylinder into which the segment is emitting is $2\pi \sigma d^2$, where $d$ is the distance from the segment where the measurement is taken. So, given that flux is luminosity per unit area, we can write the flux observed at distance $d$ as
\begin{equation}
    F_\nu=\frac{L_\nu}{2\pi \sigma d^2 }
\end{equation}
using the small-angle approximation to calculate the cylinder height as $\sigma d$.

Because the emission is directional and not isotropic, we must also consider the angle between observer and emission direction and determine how much, if any, of the emission is visible at a given viewing angle. We can write the fraction of emission that reaches the observer from a particular segment of the flux tube as
\begin{equation}
    f_{j,obs} = e^{-\Delta\theta_j^2/(2\sigma^2)}
\end{equation}
where $\Delta\theta_j$ is the angle between the segment's magnetic field $\mathbf{B_j}$ and the plane of the sky, and $\sigma$ is the thickness of the emission cone. When the observer is looking down the flux tube segment (parallel to $\mathbf{B_j}$, $\theta=\pi/2$), they will not see ECM emission, while when they are positioned side-on (orthogonal to $\mathbf{B_j}$, $\theta=0$) they will see the maximum emission (this fomulation is also used in \citet{Llama_2018ApJ...854....7L}). 

Putting this all together, we calculate the observed ECM flux from a magnetic field segment (if conditions allow emission) as
\begin{equation}
    F_{j}=e^{\frac{-\Delta \theta_j^2}{2 \delta^2}} \frac{E_{tot}/(\tau\Delta \nu )}{2 \pi \sigma  d^2 } \frac{d s_j \Omega_j}{\sum\limits_{j} d s_j \Omega_j}
\end{equation}
where all the symbols are as defined above.

\subsubsection{Dynamic spectrum creation}

\begin{figure}
\centering
	\includegraphics{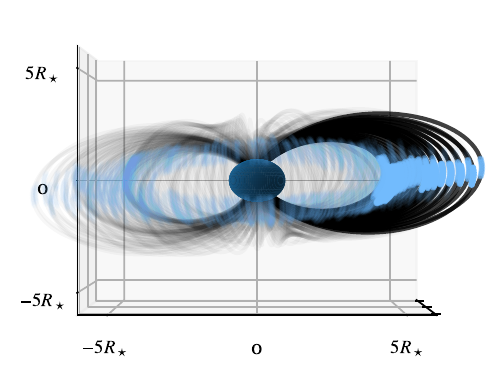}
    \caption[\peggy ejected prominence selection]{Example selection of prominences to be ejected on \peggy. The criterion used to select the ejected prominences is that the longitude $\phi$ lies in the range $80\degree<\phi<100\degree$.}
    \label{fig:ejected_prom}
\end{figure}

To create a synthetic dynamic spectrum, the first step is to choose the prominence(s) to eject. Our model is a collection of distinct magnetic field lines, however, a real corona is more continuous, so we select a bundle of closely spaced field lines and eject their prominences together, to simulate a real prominence of a specific mass being ejected (Fig. \ref{fig:ejected_prom}). We then calculate the ECM flux along each of the field lines with an ejected prominence using the method outlined above. Because of the directionality of ECM  emission, the flux observed depends heavily on the observing angle. So, for the dynamic spectrum, we calculate the flux across the observation frequency range for viewing angles throughout an entire stellar rotation. This is essentially freezing the ECM emission in place and rotating the star, as this emission is not steady-state and will only last until the ejected prominence energy is depleted. Thus, the resulting dynamic spectra is what would be observed if the dissipation time of this emission was at least as long as a stellar rotation period. Additionally, because this model assumes that ECM is turned on instantly and remains static through the dissipation time, it cannot capture frequency drifts associated with the electron motion along the field lines (for example as in \citet{Zarka_2025A&A...695A..95Z}).

\begin{figure}
\centering
	\includegraphics{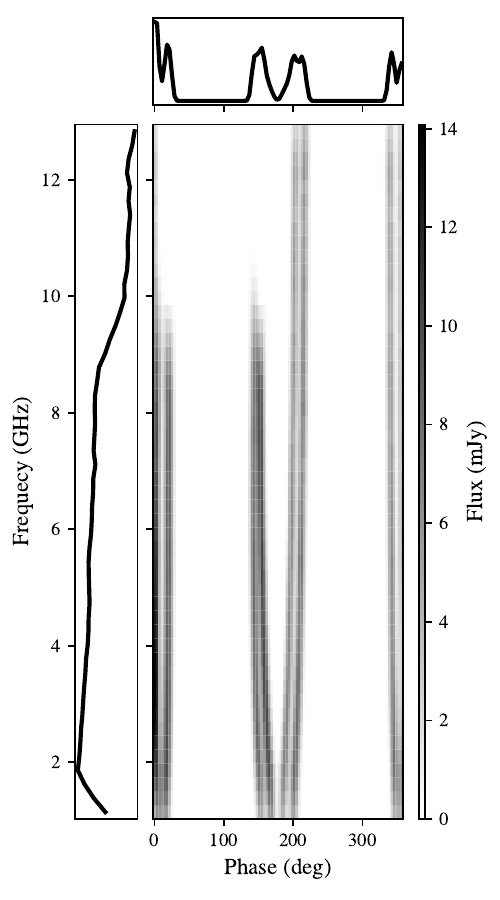}
    \caption[\peggy dynamic spectrum]{\peggy dynamic spectrum resulting from ejecting the prominences highlighted in Fig. \ref{fig:ejected_prom} {and calculating the third harmonic ECM emission}. The associated light curve is shown across the top, and the SED on the side. Note that while the instantaneous flux values can be quite high, because of resolution limits, the observed flux from a real telescope will be much lower.}
    \label{fig:dyn_spec}
\end{figure}

Fig. \ref{fig:dyn_spec} shows the dynamic spectrum resulting from ejecting the prominences highlighted in Fig. \ref{fig:ejected_prom}. In this dynamic spectrum, we can see that there are two phases where we see flux across the frequency range, and two more phases where we see flux at only a subset of the emission frequencies. The rest of the time, no flux is observable due to the directional nature of the emission combined with the viewing angle. The light curve plot (\textit{top}) shows four peaks. The reason that there are four peaks rather than two, and that the frequencies observed vary between them, is that the rotation axis is inclined to the observer by $\sim70\degree$, and the dipole axis is inclined $13\degree$ from that. Thus, the observer is not entirely edge on to the magnetic field, and therefore there is not total symmetry as we observe the star, even though the field itself is quite symmetric.

\subsection{Effect of source surface radius}

\begin{figure}
	\includegraphics{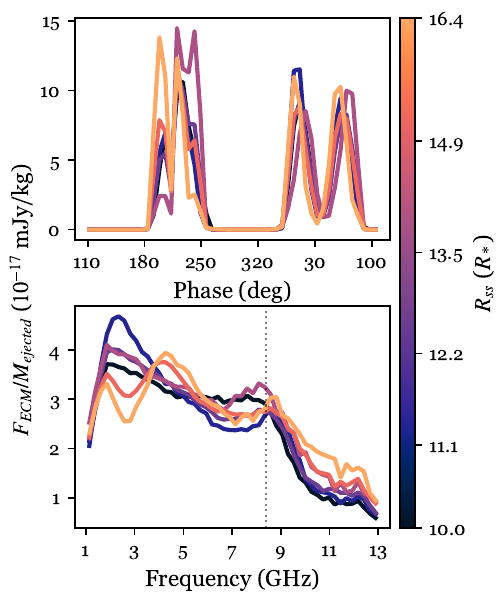}
    \caption{The effect on \peggy's ECM emission of varying only the source surface radius ($R_{ss}$) and holding base pressure and temperature constant. The \textit{top} plot shows this in phase space (light curve) and the \textit{bottom} in frequency space (SED). {The frequency observed in \citet{Hallinan_2009AIPC.1094..146H} (8.4 GHz) is marked on the lower plot}. The fluxes are normalised by ejected mass, so the resulting fluxes are not due to different levels of available energy.}
    \label{fig4:vpeg_rss}
\end{figure}

Before we move on to the results, in recognition of the fact we estimated the source surface radius from essentially an arbitrary flux value (because we do not know how much of the quiescent flux is attributable to free-free as opposed to ECM emission) it is important to understand how sensitive our model is to our selection of $R_{ss}$. In Fig. \ref{fig4:vpeg_rss} we vary $R_{ss}$ from $10-20~R_*$ and show the resulting ECM flux normalised for prominence mass across rotation phase and emission frequency. While the details of the lines do vary, the overall shapes remain consistent. Additionally, while there are some substantive differences in the spectra at low frequencies, {those are not the frequencies for which we have data}. Thus, we feel comfortable that the uncertainties in our selection of $R_{ss}$ will not affect the results. We do want to note, however, that increased understanding of the level of ECM flux from \peggy would place limits on its quiescent free-free flux, and thus also shed light on the size of its corona.

\subsection{Effect of stellar inclination}

\begin{figure}
	\includegraphics{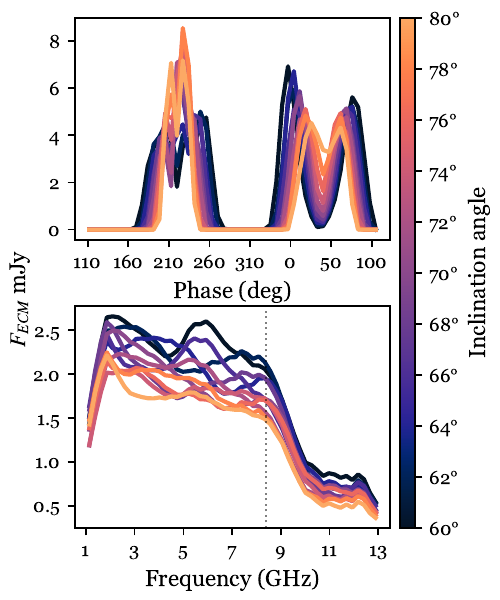}
    \caption{\peggy's inclination has an uncertainty of 20$\degree$ ($70 \pm10\degree$), so here we show the effect of varying the inclination throughout that range. The \textit{top} plot shows this in phase space (light curve) and the \textit{bottom} in frequency space (SED). {The frequency observed in \citet{Hallinan_2009AIPC.1094..146H} (8.4 GHz) is marked on the lower plot}.}
    \label{fig4:vpeg_inc}
\end{figure}

Here we explore the dependence of our results on the stellar inclination of \peggy. \peggy's inclination is known with an uncertainty of $\pm10\degree$, so we consider inclinations between $60-80\degree$. What we find is that our results are only weakly dependent on stellar inclination. As we can see in {Fig. \ref{fig4:vpeg_inc},} changing the stellar inclination changes the shape of the light curve, making each of the two broad emission peaks more or less of a doublet, while the SED remains nearly consistent in shape with small changes in overall flux level. These are entirely geometric effects, due to the change in angle between observer and magnetic field lines, making slightly different flux from different segments of field lines visible. We can see from Fig. 10 that a change in inclination changes the details of the visible flux, but the larger properties of where the flux peaks in frequency and phase space remain consistent. Additionally, the total flux observed varies much less than the uncertainties in amount of flux produced through the ECM instability.

\section{Results and Discussion}

\citet{Hallinan_2009AIPC.1094..146H} observed \peggy using the VLA in 2005. They observed for 12 hours on three consecutive nights in the VLA X-band (8-12 GHz), {specifically at $8.4 \pm 0.05$ GHz} and observed 3 clear bursts, with an additional marginal detection. The bursts they detect range in flux from $\sim0.25-5.75$ mJy, and three of the four detections occur at the same phase in the star's rotation.

The fact that \citet{Hallinan_2009AIPC.1094..146H} do not see bursts with half-phase differences indicates that {for ECM emission to be the mechanism,} the ejection energy dissipation time must be less than half a the {star's rotation period.} This is in contrast to the synthetic {dynamic spectra} shown in this paper, where the emission is assumed to last through an entire stellar rotation. {Consequently, when creating synthetic light curves, we choose a specific phase at which the ejection occurs, and only use the portion of the synthetic dynamic spectrum from the phase of ejection to the end of our chosen dissipation time.}

\subsection{{Magnetic field strength and emission frequency range}}

As discussed in \citet{Llama_2018ApJ...854....7L}, the \peggy\ VLA data were taken in the X band (8–12 GHz)\footnote{{Note that although \citet{Llama_2018ApJ...854....7L} correctly identified the VLA configuration as the X band, they misstated its frequency range as 4–8 GHz; the correct range is 8–12 GHz}.}. {This is the reason the spectrum shown in Fig.~\ref{fig:dyn_spec} was generated for the third-harmonic ECM emission rather than the fundamental.}

\begin{figure}
\centering
	\includegraphics{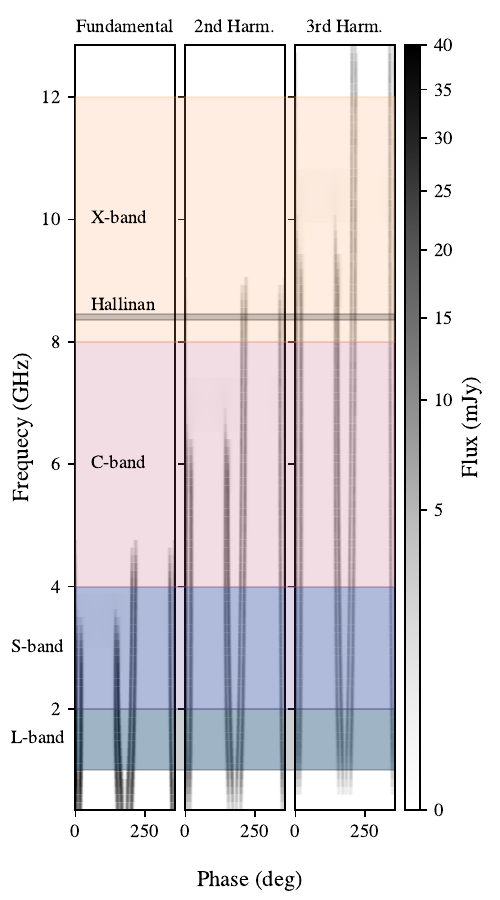}
    \caption{\peggy dynamic spectra for the fundamental frequency and first two harmonics. The prominence ejected is shown in  Fig. \ref{fig:ejected_prom}. The VLA observing bands are overlaid, {as well as the \citet{Hallinan_2009AIPC.1094..146H} observing range.}}
    \label{fig4:vpeg_harms}
\end{figure}

Fig. \ref{fig4:vpeg_harms} shows dynamic spectra at the fundamental frequency and first two harmonics, as a result of ejecting the prominence shown in Fig. \ref{fig:ejected_prom}. Note that in this plot, the same amount of energy is assumed to be available at each harmonic. This is because we do not have any information about the relative strength of the emission in each harmonic, and while it might seem reasonable to assume decreasing power at higher harmonics, this trend is not always observed \citep{Villadsen_2019ApJ...871..214V}. {Moreover, as discussed below, uncertainties in the overall magnetic-field strength introduce substantially more uncertainty into the base power levels than any reasonable treatment of the harmonic-power distribution.}

{A well-known limitation of ZDI is that it recovers only the large-scale magnetic field. ZDI maps are resolution-limited: if regions of opposite polarity lie too close together, their spectral signatures are not sufficiently separated by rotational Doppler broadening, causing them to cancel out. In addition, ZDI inversion is an ill-posed problem that requires regularization; the map used here employs a maximum-entropy constraint, which minimizes the number of spherical-harmonic modes required to reproduce the Stokes~V profiles and therefore favours simpler field geometries. Previous work shows that ZDI reliably captures the large-scale component of the field, which is most relevant for radio emission. However, X-ray studies indicate that a substantial fraction of the magnetic energy resides in small-scale fields. Estimates in the literature suggest that ZDI may miss a significant portion of the total magnetic flux, implying that the true field strengths could be considerably higher \citep{Lang_2014MNRAS.439.2122L,Hahlin_2023A&A...675A..91H,Hackman_2024A&A...682A.156H, Yadav_2015ApJ...813L..31Y}. For \peggy \citet{Bellotti_2025arXiv251109312B} suggest a small-scale field strength of $\sim5.5$ kG. This, in turn, would shift the expected ECM frequencies upward, easily encompassing the observed frequency range.}

{As seen in Fig.~\ref{fig4:vpeg_harms}, the emission spans a broader frequency range with each successive harmonic. By contrast, increasing the total magnetic energy would shift the fundamental distribution to higher frequencies. In principle, high-fidelity broadband spectra could allow these two scenarios to be distinguished observationally, but with the existing data this is not possible. For this work we therefore adopt the conservative assumption that the observed emission corresponds to the third harmonic, without invoking additional unresolved magnetic flux.}

\subsection{ECM Flux Range}

\begin{figure}
\centering
	\includegraphics{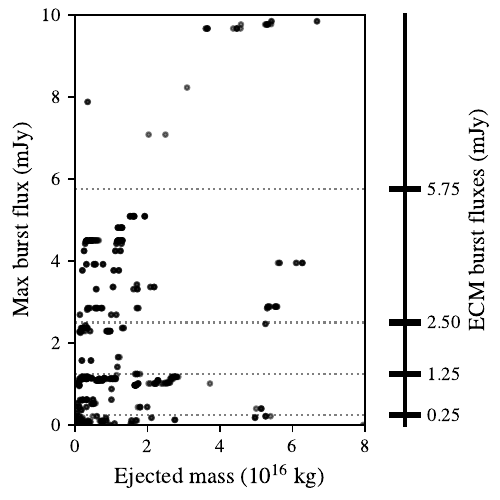}
    \caption[\peggy ejected mass vs. ECM flux]{Ejected mass vs. ECM radio burst flux on \peggy. The ECM flux is from the second harmonic calculated over the VLA X-band range, and model parameters are given in Table \ref{tbl:vpeg_model}.}
    \label{fig4:eje_mas_flux}
\end{figure}

In the previous sections, we have shown the results of ejecting prominences from a specific selection of prominence lines, however, there are any number of combinations of magnetic field lines that can eject prominences. Fig. \ref{fig4:eje_mas_flux} shows the range of ECM flux values in the VLA X-band attainable with the range of ejected prominence masses. 

We determined the range of possible ejection masses by selecting all the prominences within an angular range between $0\degree$ (a single prominence bearing line) and $30\degree$, summing the individual prominence masses to get the total ejected mass, and calculating the result of ejecting all selected prominences at once. We repeated this selection process, starting at 30 evenly spaced values of $\phi$ around the star. We also capped the prominence mass at $8\times10^{16}$ kg, because beyond that mass, the ECM fluxes were unreasonably high compared to observed values. The number bar at the right of the plot shows the observed radio burst fluxes from \citet{Hallinan_2009AIPC.1094..146H}. We note particularly that the range of fluxes attained by our model easily encompasses the range of observed values.

At this point, we want to emphasise that the precise range of ejected mass is not our central concern, given the significant uncertainties involved. These uncertainties primarily include the strength of the second harmonic and the efficiency factor, both of which are not well constrained beyond a general sense of plausibility. However, the key argument is that, within a reasonable range of parameters and ejected masses, this mechanism can readily account for the observed flux values.

\subsection{Burst Timings}

\begin{figure*}
\centering
	\includegraphics{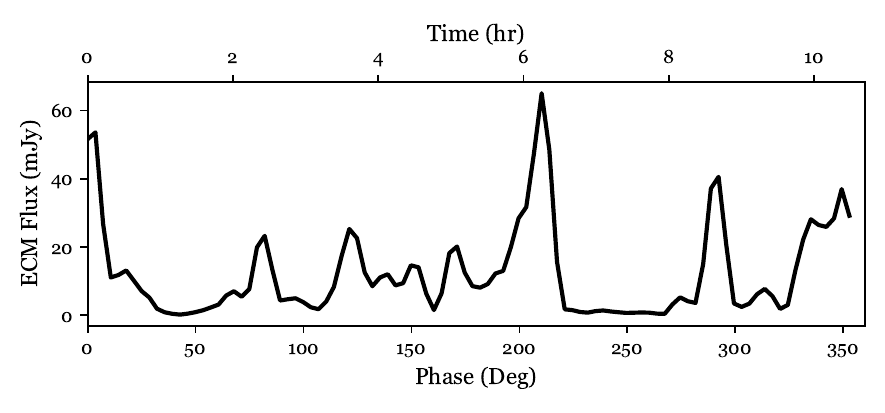}
    \caption{This plot shows the result of ejecting every prominence at once and assuming the dissipation time is more than a stellar rotation, so that all of the emission is present at once and for a full rotation. In this plot, we can see that certain phases are favoured for ECM emission. This plot was made using the {third} harmonic and calculated over the {\citet{Hallinan_2009AIPC.1094..146H} frequency} range.}
    \label{fig4:all_ejections}
\end{figure*}

Having now calculated the range of ECM burst energies that can be produced, we turn to the timing. \citet{Hallinan_2009AIPC.1094..146H} note that three of the four bursts they observe occur at the same rotational phase, so to see if there is a phase preference in our synthetic ECM data, we eject all of the prominences at once and plot the result in Fig. \ref{fig4:all_ejections}. In this plot, we can see that there is indeed a phase preference, and in fact there are two maxima about 2 hours apart, which is the phase difference between the \citet{Hallinan_2009AIPC.1094..146H}'s fourth observed burst and the other three. These phase preferences correspond to regions of the magnetic field which can support more prominence mass, and hence generate more ECM emission.

\subsection{Example Light Curves}

\begin{figure}
\centering
	\includegraphics{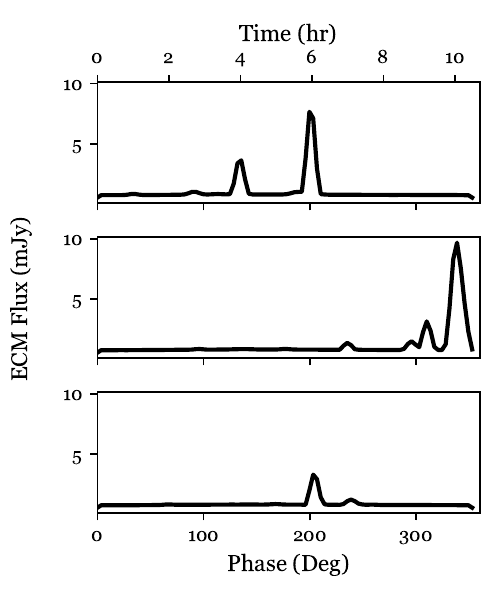}
    \caption[Example \peggy light curves with ECM and free-free emission]{This plot shows a selection of example light curves, the \textit{top} with two, \textit{middle} with three, and \textit{bottom} with four prominence ejections inserted at randomly selected phases (thus not all of the ejections will be visible). The dissipation time for the ECM bursts is set to 6 hours. Free-free emission is also included in these light curves. Note the variety of morphology and flux range, as well as the visible preference for certain phases.}
    \label{fig4:example_lcs}
\end{figure}

Finally, we combine ECM flux with the free-free light curve, calculated using the methods described in \citet{Brasseur_2024MNRAS.530.2442B} to build a number of example light curves. The way we build the light curves is by selecting a number of prominences to eject (which individual prominences are selected, and their masses, are random), and then choosing a random set of phases at which the ejections occur. We also choose a dissipation time so that the ejection-induced flux will only be visible over that dissipation time. For simplicity, we use the same dissipation time of 6 hours for all ejections, although in principle it could be scaled in some way. We choose this dissipation time because it is just under half a rotation of the star, so it is consistent with \citet{Hallinan_2009AIPC.1094..146H}'s observations. We then combine the ECM light curve from each ejection and the free-free light curve. Because of the highly directional nature of the ECM emission, depending on the phase at which a prominence is ejected, it may never be visible in the light curve. 

Fig. \ref{fig4:example_lcs} shows a selection of these synthetic light curves, all of them made by ejecting between two and four prominences at randomly selected times. There are a few features we want to direct the reader's attention to. Firstly, the varied characteristics of the light curves; a variety of burst timings and fluxes are on display, however, the preference for certain phases is also visible. Secondly, the sinusoidal quiescent flux variation {described by \citet{Hallinan_2009AIPC.1094..146H}} is not visible. As mentioned in \S\ref{ch4:vpeg_find_parms} with the parameters we use in this model (specifically the low base densities), {the free-free flux variation, while sinusoidal, has very low amplitude (on the order of 1/20 mJy)}. Thus, there must be another source of emission driving {the quiescent} flux, such as the quiescent ECM emission theorised by \citet{Hallinan_2009AIPC.1094..146H} and modelled by \citet{Llama_2018ApJ...854....7L}. {Lastly, the widths of the synthetic bursts match the observed widths, all on the order of one hour.}

\section{Conclusions}

In this study, we have modelled transient ECM emission from ejected stellar prominences and shown that it is a viable mechanism for achieving the observed radio bursts in the \peggy light curve. In this scenario, as slingshot prominences are ejected (as is observed on \peggy when prominences become too massive to be supported by the magnetic field), the resulting kinetic energy drives a magnetic reconnection event, which in turn induces ECM emission, and that ECM emission can be observed in the form of radio bursts. Additionally, we have shown the following about ECM emission powered by stellar prominence ejection:

\begin{itemize}
    \item It reproduces the observed preference for certain rotation phases.
    \item It reproduces the range of observed radio burst fluxes.
    \item Higher-order harmonics contribute to the observed emission.
    \item Ejection-induced ECM dissipation time must be less than half the stellar rotation period ($\lesssim 6$ hr).
    \item To reproduce the background emission, it must be combined not only with free-free emission, but also with some other emission mechanism, such as quiescent ECM emission as modelled by \citet{Llama_2018ApJ...854....7L}.
\end{itemize}

This study demonstrates the power of coupling observation-driven modelling with independent observations. This work also points the way toward further such studies of other M Dwarfs to explore the conditions under which this type of ECM emission is and is not observable.

\section*{Acknowledgements}

This work has made use of data from the European Space Agency (ESA) mission
{\it Gaia} (\url{https://www.cosmos.esa.int/gaia}), processed by the {\it Gaia}
Data Processing and Analysis Consortium (DPAC,
\url{https://www.cosmos.esa.int/web/gaia/dpac/consortium}). Funding for the DPAC
has been provided by national institutions, in particular the institutions
participating in the {\it Gaia} Multilateral Agreement.

MMJ acknowledges support from STFC consolidated grant number ST/R000824/1. CEB acknowledges funding from the St Andrews GH Chaplin scholarship. The authors have applied a creative commons attribution (CC BY) licence to any author accepted manuscript version arising.

\section*{Data Availability}

Please contact the authors to request data access. The software used in this work can be accessed at \url{https://coronalab.readthedocs.io}.



\bibliographystyle{mnras}
\bibliography{StellarAtmospheres,Brasseur,StellarProperties,V374Peg,Prominences,Solar,StellarTechniques,Reference,StarsAndPlanets}  





\bsp	
\label{lastpage}
\end{document}